\documentclass{JHEP3}
\usepackage{amsmath}

\title{On the universality of hadronisation corrections to QCD jets}

\author{Mrinal Dasgupta\\
        School of Physics and Astronomy, University of Manchester\\
        Oxford road, Manchester M13 9PL, U.K.\\
        \email{Mrinal.Dasgupta@manchester.ac.uk}}

\author{Yazid Delenda\\
        D\'epartement de Physique, Facult\'e des Sciences\\
        Universit\'e de Batna, Algeria.\\
        \email{yazid.delenda@yahoo.com}}

\preprint{MAN/HEP/2009/12}

\abstract{We improve  previously derived analytical estimates of
hadronisation corrections to QCD jets at hadron colliders, firmly
establishing at the two-loop level the link to the well-known power
corrections to LEP event-shape variables. The results of this paper
apply to jets defined in the $k_t$ and anti-$k_t$ algorithms but the
general framework presented here holds also for other algorithms for
which calculations are in progress.}

\keywords{QCD, Jets}

\begin{document}

\section{Introduction}

The study of processes involving the production of high transverse
momentum ($p_t$) QCD jets will form an integral part of the LHC
physics programme. In this light it is clear that the understanding
of properties of jets will impact the accuracy of studies concerning
new physics as well as traditional QCD studies concerning for
instance the extraction of parton distribution functions.

While the subject of jet physics is far from new and has seen steady
progress since the pioneering work of Sterman and Weinberg
\cite{SW}, the LHC provides a challenge of unprecedented complexity
owing to the initial state hadronic environment and the very high
beam energy. The QCD tools at our disposal have therefore to be
developed and adapted to meet this challenge.

The most important tool in the context of accurate jet studies at
the LHC will remain QCD perturbation theory. Reliable perturbative
estimates will constitute the only first principles
(model-independent) information available from QCD theory and hence
accurate perturbative results for various processes of interest are
imperative. The advent of new infrared and collinear (IRC)-safe jet
algorithms such as SISCone \cite{SISCONE} and the anti-$k_t$
\cite{antikt} jet algorithms to complement  existing IRC-safe
algorithms such as the $k_t$ \cite{inclukt,ESkt} and the
Cambridge/Aachen algorithms \cite{Caachen} is therefore an
encouraging and crucial development.

Perturbative aspects apart, another serious stumbling block for jet
studies at the LHC is the role of non-perturbative effects. Here one
is dealing not just with the hadronisation of partons but also the
underlying event arising from beam remnant interactions, multiple
hard interactions and the issue of pile-up. Thus while in an ideal
world one may hope to construct clean signals for new physics, for
instance mass peaks signifying the production of new particles, in
the real hadron collider environment such peaks can be significantly
smeared by non-perturbative QCD effects such as hadronisation and
the underlying event, in addition to perturbative radiation. A
better understanding of the size and role of non-perturbative
effects as a function of the parameters used in jet studies (such as
jet size) is hence imperative.

While neither hadronisation nor the underlying event can be
calculated within perturbation theory one could argue that one may
have more control on the former effect. The reason for this argument
is that the hadronisation of partons that constitute a high-$p_t$
jet is a phenomenon that has been met and handled before, for
instance at LEP, HERA and the Tevatron. There has therefore been
considerable development of hadronisation models in Monte Carlo
event generators such as {\tt PYTHIA} \cite{pythia} and {\tt HERWIG}
\cite{herwig}.

Also significantly, in addition to Monte Carlo simulations,
analytical techniques based on renormalons \cite{Beneke} have
yielded important information on hadronisation corrections in the
context of event-shape variables at LEP and HERA \cite{Dassalrev}.
The pioneering phenomenological studies in this regard were carried
out by Dokshitzer and Webber \cite{DokWeb}. Using a model for a
universal infrared-finite coupling they were able to compute the
$1/Q$ power corrections to event-shape variables at LEP. The ensuing
predictions involved just a single universal non-perturbative
parameter, the moment of the coupling over the infrared region.
Extracting the value of this moment from data on one variable it
thus became possible to predict the $1/Q$ corrections to a large
class of event-shape variables. Similar success was also met in the
description of Breit-frame current-hemisphere event-shape variables
at HERA \cite{Dassalshapes} in spite of the presence of an
initial-state proton that may potentially have impacted the power
corrections.

The study of hadronisation corrections at hadron colliders, based on
analytical techniques, is however still in its infancy. The main
deterrents to such studies have been for instance the more
complicated colour topology of multijet hard processes at hadron
colliders and the competing presence of the underlying event. The
issue of the role of colour in influencing the hadronisation
corrections was explained in ref.~\cite{Dasdelhad} in the context of
away-from--jet energy flow.

As far as the role of the underlying event compared to hadronisation
is concerned an interesting result was derived in
ref.~\cite{Dassalmag}. There it was observed that if one studies the
hadronisation contribution to the jet energy (specifically the
change in $p_t$, $\delta p_t$ of a jet due to hadronisation) using
the aforementioned analytical techniques, one obtains a singular
dependence\footnote{A similar result was obtained by Korchemsky and
Sterman for cone jets in $e^{+}e^{-}$ annihilation (see ref.
\cite{KS}).} on the jet radius $R$ with the hadronisation varying as
$-1/R$. The underlying event on the other hand makes a contribution
to the jet $p_t$ varying as $R^2$ (see also refs.~\cite{CS,CSS}).
The possibility thus emerges of a parametric separation of the two
components of non-perturbative QCD with hadronisation being the
dominant effect for jets at small $R$. Moreover the leading
small-$R$ behaviour was seen to arise due to \emph{collinear} soft
gluon radiation from the triggered hard parton jet. The colour
factor associated to collinear radiation is merely the colour charge
of the emitting hard parton, which means that in this limit the
complex colour topology of the whole event is essentially
irrelevant. Knowing the $R$-dependence of perturbative and
non-perturbative corrections then allows for the possibility of
deducing optimal $R$ values adapted to different studies at hadron
colliders \cite{Dassalmag}.

Perhaps most interestingly a link was also made in
ref.~\cite{Dassalmag} between the magnitude of the $1/R$
hadronisation correction to the jet $p_t$ and that of $1/Q$ power
corrections to event shapes such as the thrust in $e^{+}e^{-}$
annihilation \cite{DokWeb}. The coefficient of the $1/R$
contribution is a perturbatively calculable number times a
non-perturbative factor which is the same coupling moment as enters
the event shape predictions. Carrying out the relevant perturbative
calculation at the single-gluon level \cite{Dassalmag} the
coefficient of the $1/R$ correction to the jet $p_t$ for a quark jet
is seen to be one-half of the coefficient of the $1/Q$ power
correction for the $e^{+}e^{-}$ thrust variable as computed in
ref.~\cite{DokWeb}.

The predictions of refs.~\cite{DokWeb} and \cite{Dassalmag} are
based on one-loop ${\mathcal{O}}(\alpha_s)$ perturbative
calculations. At this level the emission of a soft gluon with
transverse momentum $k_t \sim \Lambda_{\mathrm{QCD}}$ is associated
to non-perturbative hadronisation contributions to the transverse
momentum $p_t$ of a hard jet or the event shape under consideration.
Physically this is due to the fact that the running coupling
associated to such an emission is $\alpha_s(k_t)$ which is
perturbatively undefined at $k_t \sim \Lambda_{\mathrm{QCD}}$.
Replacing the unphysical perturbative coupling in this region by a
finite and universal non-perturbative extension\footnote{The
treatment of the running coupling was formalised via the dispersive
approach adopted in ref.~\cite{DMW}.} leads to the predictions for
hadronisation corrections reported in refs.~\cite{DokWeb} and
\cite{Dassalmag}.

In the context of event shapes however it was shown by Nason and
Seymour that one-loop calculations involving single gluon emission
are inadequate to predict the coefficient of the $1/Q$ power
corrections \cite{NasSey}. The argument of the running coupling
which we took to be $k_t$ can in fact only be reconstructed at the
two-loop level where one has to account for the decay of the emitted
gluon into offspring gluons or a quark/anti-quark pair. For
variables such as event shapes and some observables involving jet
definition such as jet $p_t$, one is \emph{not} free to inclusively
integrate over the decay products since the observable depends on
the precise details of the decay kinematics. Thus in these cases the
replacement of a gluon decay with a parent virtual (massive) gluon,
with a running coupling dependent on the parent virtuality or $k_t$,
is not actually correct at the level of power corrections. One must
then address the gluon decay and identify the correction needed to
the single gluon results quoted for example in refs.~\cite{DokWeb}
and \cite{Dassalmag}.

For the case of the thrust and other commonly studied $e^{+}e^{-}$
event-shape variables calculations accounting for gluon decay have
already been performed some time ago \cite{Milan1, Milan2} and
yielded a simple result: taking into account gluon decay resulted in
a universal (event-shape--independent) factor multiplying the single
massive gluon results -- the ``Milan'' factor. This was also seen to
be the case for DIS event shapes \cite{MilanDW} and further
confirmed by the results of ref.~\cite{MilanDMS}. One of the key
ingredients in obtaining a universal Milan factor was the fact that
all the event-shape variables considered are \emph{linear} in
transverse momenta of soft emitted particles \cite{Milan2}. The jet
$p_t$ variable is however not so simple. For almost all the commonly
used jet algorithms it turns out that the $\delta p_t$ due to gluon
emission is not simply a linear sum over the contributions of
individual gluons, and that the contribution of a given gluon
depends on whether or not it is clustered or combined with other
emissions. This effect is absent for the anti-$k_t$ algorithm where
soft gluons essentially cluster independently to the hard emitting
parton and the result is the universal Milan factor \cite{antikt}.

In the present paper we establish that for jets defined in, for
instance, the $k_t$ algorithm (which we treat explicitly) the
resulting rescaling factor is \emph{not} the universal Milan factor.
However we prove that it is a calculable factor of the same order,
which we compute for the $k_t$ algorithm. Thus it is now indeed
possible to link the results for hadronisation corrections to jet
$p_t$ with those obtained for event shapes after including the Milan
factor. As a result of our calculations the relation quoted in
ref.~\cite{Dassalmag} between the jet $p_t$ hadronisation and the
thrust power correction changes for all but the anti-$k_t$
algorithm. The hadronisation corrections to jet $p_t$ become
algorithm-dependent rather than truly universal. They however remain
universal in the more crucial sense that the result is always a
perturbatively calculable number multiplying a universal
non-perturbative coefficient. The only complication is that the
perturbative calculations have to be at the two-loop level.

The current paper is organised as follows. In section
\ref{sec:kinematics} we deal with the impact of gluon emission and
decay on the jet $p_t$ considering both the kinematical effect on
the observable as well as the dynamics of gluon branching. The
na\"{i}ve one-gluon term calculated in ref.~\cite{Dassalmag} is
identified and seen to be accompanied by two distinct correction
terms, which following refs.~\cite{Milan1,Milan2} we refer to as the
inclusive and non-inclusive corrections. Section
\ref{sec:dispersive} deals with extracting the $1/R$ behaviour for
the na\"{i}ve result (where we recover the answer quoted in
ref.~\cite{Dassalmag}) and computing the inclusive correction.
Section \ref{sec:ni-trigger} is devoted to the treatment of the
non-inclusive term for the $k_t$ algorithm and the extraction of the
leading $1/R$ behaviour in this term. In section \ref{sec:conc} we
combine the results to derive an overall rescaling factor to the
result of ref.~\cite{Dassalmag} for the $k_t$ algorithm. We also
recover the expected result for the anti-$k_t$ algorithm which
agrees with the Milan factor found for event shapes \cite
{Milan1,Milan2}. Finally we draw some conclusions and discuss
prospects for further work.

We should also remark that in the following sections we make rather
heavy use of notation and terminology as well as many results common
to refs.~\cite{Milan1, Milan2, MilanDW}, since most aspects of the
calculation here carry over unaltered from those papers. Thus a full
technical understanding of some ideas used in this article may
require a reading of those references. The main ideas and the
essential calculations should however be clear on a reading of just
the current paper.

\section{The jet $\delta p_t$ and gluon
splitting}\label{sec:kinematics}

Here we repeat the arguments, first detailed in ref.~\cite{Milan1},
that led to the emergence of the universal Milan factor, but do so
in the context of the jet $\delta p_t$. We start with some
introductory remarks that explain the simplifications we make which
lead us to the final results.

\subsection{Preliminary remarks}

In ref.~\cite{Dassalmag} it was noted that the dominant piece of the
non-perturbative hadronisation contribution to the jet $p_t$ at
small jet radius $R$ scaled as $1/R$. Additional terms
$\mathcal{O}(R)$ and beyond were also obtained but these had a
negligible impact at small\footnote{As a matter of fact the $1/R$
behaviour was shown to hold to a good approximation over virtually
the entire range of $R$ values up to $R=1$, which can in part be
attributed to the smaller coefficients that emerged for the finite
$R$ pieces in the analysis of \cite{Dassalmag}.} $R$. Due to its
significance we shall concentrate in this article on the dominant
$1/R$ piece of the result.

It was also made clear in ref.~\cite{Dassalmag} that the $1/R$ term
arose from gluon emission at the boundary of the jet, which in the
small-$R$ limit is in the region collinear to the hard parton
initiating the jet. To be precise the singular dependence on $R$ is
a direct manifestation of the collinear singularity present in the
splitting of massless partons due to gluon emission. The same
collinear singularity was also shown to be associated with $\ln R$
terms in the perturbative regime, which we do not discuss further
here.

Given its origin in collinear radiation it is simple to deduce that
the $1/R$ term does not depend on the details of the process of
which the triggered hard parton (jet) is a part. This fact
considerably simplifies the calculations involved and rather than
considering all possible hard emitting dipoles in a given hard
process (such as dijet production in hadron-hadron collisions
treated in refs.~\cite{Dasdelhad,Dassalmag}) one need only consider
collinear branching of the massless parton corresponding to the
triggered hard jet.

In spite of this obvious generality, to illustrate the calculation
and clarify the connection with the existing treatment
\cite{Dassalmag} it may be useful to take a specific example. The
example chosen in ref.~\cite{Dassalmag} was the change in $p_t$ due
to hadronisation, $\delta{p_t}$, of a jet produced in the threshold
limit of hadronic dijet production where threshold was defined by
the limit $p_t \to \sqrt{s}/2$. Then the change in $p_t$ from its
threshold value due to soft gluon emission was computed and the
gluon emission was averaged over to obtain the mean value of $\delta
p_t$. In what follows we shall repeat the calculation of
ref.~\cite{Dassalmag}, this time in the {\emph{collinear
approximation}} sufficient to generate the $1/R$ behaviour but with
account of the decay of the emitted gluon.

\subsection{Gluon emission kinematics and dynamics}

We shall also be making use of the formulae and notation used in the
Milan factor computations and first presented in ref.~\cite{Milan1}.
To this end let us introduce Sudakov vectors along and opposite to
the triggered jet direction, taken itself to be at ninety degrees
with respect to the beam as required by the threshold limit:
\begin{align}
P &= \frac{\sqrt{s}}{2}(1,1,0,0) ,\nonumber\\
\bar{P} &= \frac{\sqrt{s}}{2}(1,-1,0,0).
\end{align}
We parameterise the emitted parton four-momenta as in
ref.~\cite{Milan1}:
\begin{equation}
k_i= \alpha_i P +\beta_i \bar{P} +k_{ti}\,,
\end{equation}
with $\vec{k}_t$ the transverse momentum vector with respect to the
jet direction. For massless partons we have that $k_i^2 =0$ or
$\alpha_i \beta_i = k_{ti}^2/s$. Also with neglect of recoil of the
hard partons against the soft parton momenta we can identify $P$ and
$\bar{P}$ with the momenta of the final state hard partons
themselves (recall that we are considering threshold where the hard
partons are produced back-to--back and at ninety degrees to the
beam).

We shall also need the value of $\delta p_t$ due to soft parton
emissions. Here one can make use of the results derived in
ref.~\cite{Dassalmag}, where it was shown that emitted partons that
are recombined with the triggered hard parton contribute to the
$\delta p_t$ by lending a mass $M_j^2$ to the jet. Soft partons not
recombined with the jet lend a mass to the ``recoil jet'' involving
the recoiling hard parton which is not triggered. Thus one has:
\cite{Dassalmag}:
\begin{equation}
\label{eq:partin} \delta p_t = p_t-\frac{\sqrt{s}}{2} = -
\left(\frac{M_j^2}{2 \sqrt{s}}+\frac{M_r^2}{2 \sqrt{s}} \right),
\end{equation}
where the above equation is correct to first order in soft emitted
parton momenta or equivalently the small jet masses. In terms of the
Sudakov variables defined above one obtains:
\begin{equation}
\label{eq:injet} \frac{M_j^2}{2 \sqrt{s}} = \frac{1}{2\sqrt{s}}
\left( p+\sum_{i \in j} k_i \right)^2 = \frac{\sqrt{s}}{2} \sum _{i
\in j} \beta_i\,,
\end{equation}
where the sum runs over all emissions recombined with the triggered
hard parton, to form the triggered jet, and we neglected all terms
quadratic in the transverse momenta of soft emitted partons
including those that give a tiny recoil to the hard parton momentum
$p$, which we thus identified with the Sudakov vector $P$. Likewise
the mass of the recoil system is:
\begin{equation}
\label{eq:partout} \frac{M_r^2}{2 \sqrt{s}} = \frac{\sqrt{s}}{2}
\sum_{i \notin j} \alpha_i\,,
\end{equation}
where the sum is over all emissions not recombined into jet $j$. We
shall use the above results in the following subsection.

\subsection{Calculation of $\left \langle \delta p_t \right \rangle
$ with gluon decay}

Here we shall consider the change in $p_t$ derived above, together
with the squared matrix element for gluon production and decay, to
obtain the average $\delta p_t$. We consider the situation up to the
two-loop $\mathcal{O}\left (\alpha_s^2 \right)$ level and hence we
need to account for single gluon emission, virtual corrections to it
as well as correlated two parton emission. To be precise we can
write \cite{Milan1,Milan2}:
\begin{equation}
\label{eq:int} \left<\delta p_t\right> = \frac{C_F}{\pi}\int
\frac{d^2k_t}{\pi k_t^2} \frac{d\alpha}{\alpha}
\left\{\alpha_s(0)+4\pi \chi(k_t^2)\right\} \delta p_t(k)+4 C_F \int
\left(\frac{\alpha_s}{4\pi}\right)^2 d \Gamma_2 \frac{M^2}{2!}
\delta p_t(k_{1},k_{2}),
\end{equation}
where the first term on the right-hand side of the above represents
the contribution from real massless gluon emission and the virtual
correction to single emission denoted by $\chi(k_t^2)$. The second
term on the right-hand side is the contribution from correlated
two-parton emission with the squared matrix element for gluon decay
$M^2$ containing both gluon decay into a pair of gluons as well as
the Abelian contribution from gluon branching to a $q \bar{q}$ pair.
The measure $d \Gamma_2$ is the decay phase-space and $\delta
p_t(k_1,k_2)$ is merely the change in $p_t$ induced by two-parton
emission, a special case of the general situation discussed above.

We should also mention the fact that the above result is in fact a
``dressed'' two-loop result where an infinite set of higher order
perturbative graphs are implicitly embedded in the gluon propagator.
These graphs are, at least in the Abelian limit, essentially
self-energy or bubble insertions (renormalon graphs -- see
\cite{Beneke} for a review) that drive the scale of the coupling to
the gluon virtuality $k^2$, hence triggering non-perturbative
contributions from the infrared region $k^2 \sim
\Lambda_{\mathrm{QCD}}^2$. Since for  a real gluon the virtuality
$k^2$ is zero, one has the ill-defined quantity $\alpha_s (0)$
appearing above (which will eventually disappear), while in the
gluon decay case the argument of $\alpha_s$ is $k^2=m^2$.

Note also that in writing eq.~\eqref{eq:int} we have specialised to
the collinear limit by ignoring the fact that the soft gluon $k$ can
be emitted by several dipoles that form the hard particle ensemble,
since in the collinear limit the emission is essentially off the
triggered hard parton alone. We have taken this parton to be a quark
and hence the appearance above of the colour factor $C_F$, but for a
gluon jet this can straightforwardly be replaced by $C_A$.

We now follow all the arguments of refs.~\cite{Milan1, Milan2}. We
first attempt to treat $\delta p_t$ as an inclusive quantity which
corresponds to the treatment in ref.~\cite{Dassalmag}. We then
identify that there is a non-vanishing correction needed to the
inclusive treatment which we shall compute. To this end we write:
\begin{equation}
\label{eq:diff} \delta p_t (k_1,k_2) = \delta p_t(k_1,k_2)-\delta
p_t(k_1+k_2)+\delta p_t(k_1+k_2),
\end{equation}
where the term $\delta p_t(k_1+k_2)$ corresponds to an inclusive
treatment where the true value of $\delta p_t(k_1,k_2)$ is replaced
by the value of $\delta p_t$ that would be obtained by considering
just the emission of the massive parent gluon $k_1+k_2$, with
$(k_1+k_2)^2 = m^2$. The remainder $\delta p_t(k_1,k_2)-\delta
p_t(k_1+k_2)$ leads to a correction term that we evaluate in due
course\footnote{It should be easy to see that this correction term
is in general non-zero. If one considers for example a gluon decay
where only one of the decay products say $k_1$ gets recombined with
the jet, then one has $\delta p_t(k_1,k_2) =-\sqrt{s}/2 \,
(\alpha_2+\beta_1)$. Assuming that the parent gluon is outside the
jet the inclusive contribution is $\delta p_t(k_1+k_2) = -\sqrt{s}/2
\, \alpha =-\sqrt{s}/2 \, (\alpha_1+\alpha_2)$ which differs from
$\delta p_t (k_1,k_2)$.}.

Substituting \eqref{eq:diff} into \eqref{eq:int} we can write:
\begin{multline}
\label{eq:int1} \left<\delta p_t\right> = \frac{C_F}{\pi}\int
\frac{d^2k_t}{\pi k_t^2} \frac{d\alpha}{\alpha}
\left\{\alpha_s(0)+4\pi \chi(k_t^2)\right\} \delta p_t(k)+\\+4 C_F
\int \left(\frac{\alpha_s}{4\pi}\right)^2 d \Gamma_2 \frac{M^2}{2!}
\delta p_t(k_{1}+k_2)+\left\langle \delta p_t \right \rangle^{ni}
\,,
\end{multline}
where we separated the non-inclusive correction (denoted $ni$) from
the inclusive approximation such that:
\begin{equation}
\label{eq:ni} \left\langle \delta p_t \right\rangle^{ni} = 4 C_F
\int \left(\frac{\alpha_s}{4\pi}\right)^2 d \Gamma_2 \frac{M^2}{2!}
\left(\delta p_t(k_1,k_2)-\delta p_t(k_1+k_2)\right).
\end{equation}

For the subsequent discussion we will need the results below for the
two-parton (gluon decay) phase-space \cite{Milan1, Milan2}:
\begin{equation}
d \Gamma_2(k_1,k_2) = \prod_{i=1}^{2} \frac{d \alpha_i}{\alpha_i}
\frac{d^2 k_{ti}}{\pi} = \frac{d\alpha}{\alpha} \frac{d^2k_t}
{\pi}\frac{d^2 q}{\pi} z(1-z) dz\,,
\end{equation}
with $\alpha=\alpha_1+\alpha_2$ the Sudakov variable defined
earlier, $\vec{k}_t = \vec{k}_{t1}+\vec{k}_{t2} $ the transverse
momentum of the massive parent gluon, $z$ the fraction
$\alpha_1/\alpha$ and $q$ the relative ``transverse angle'' of the
pair. Thus we have:
\begin{equation}
\alpha_1 = z \alpha\,, \qquad \alpha_2 =(1-z) \alpha\,, \qquad
\vec{k}_t = \vec{k}_{t1}+\vec{k}_{t2}\,, \qquad \vec{q}
=\frac{\vec{k}_{t1}}{z}-\frac{\vec{k}_{t2}}{1-z}\,.
\end{equation}

The above can also be straightforwardly expressed in terms of the
mass of the parent gluon $m^2 = z(1-z) q^2$ so that:
\begin{equation}
d \Gamma_2(k_1,k_2) = \prod_{i=1}^{2} \frac{d \alpha_i}{\alpha_i}
\frac{d^2 k_{ti}}{\pi} = dm^2 \frac{d\alpha}{\alpha}
\frac{d^2k_t}{\pi}dz \frac{d\phi}{2\pi}\,,
\end{equation}
where $\phi$ is the angle between $\vec{k}_t$ and $\vec{q}$.

We note that the quantity $M^2$ is invariant under boosts in the
longitudinal direction implying that it is independent of the
variable $\alpha$, which allows us to perform the $\alpha$ integral
explicitly. Defining the result of the $\alpha$ integral by a
{\emph{trigger function}} $\Omega$ we can write the inclusive piece
(first two terms on the right-hand side of eq.~\eqref{eq:int1}) as:
\begin{multline}
\label{eq:inttrig} \left<\delta p_t\right>^{i} = \frac{C_F}{\pi}\int
\frac{d^2k_t}{\pi k_t^2} \left\{\alpha_s(0)+4\pi \chi(k_t^2)\right\}
\Omega_0(k_t^2) +\\+ 4 C_F \int \left(\frac{\alpha_s}{4\pi}\right)^2
dm^2 \frac{d^2k_t}{\pi} dz \frac{d\phi}{2\pi} \frac{M^2}{2!}
\Omega_0(k_t^2+m^2),
\end{multline}
with the trigger function being:
\begin{equation}
\Omega_0 (k_t^2+m^2) = \int_{(k_t^2+m^2)/s}^1 \frac{d\alpha}{\alpha}
\delta p_t \left(\alpha,\beta=(k_t^2+m^2)/(s \alpha) \right),
\end{equation}
and similarly for $\Omega_0(k_t^2)$ which just corresponds to real
massless gluon emission, and the superfix $i$ pertains to the
inclusive piece. In writing the above we have made use of the fact
that $\delta p_t$ depends on the $\alpha$ and $\beta$ Sudakov
variables and that $\beta = (k_t^2+m^2)/(\alpha s)$ in the massive
case and $k_t^2/(\alpha s)$ for massless real emission.

For the non-inclusive contribution a trigger function $\Omega_{ni}$
can be similarly defined so that it reads:
\begin{equation}
\label{eq:ninctrig} \left \langle \delta p_t \right \rangle^{ni} = 4
C_F \int \left(\frac{\alpha_s}{4\pi}\right)^2 dm^2
\frac{d^2k_t}{\pi} dz \frac{d\phi}{2\pi}\frac{M^2}{2!}
\Omega_{ni}\,,
\end{equation}
with
\begin{equation}
\label{eq:nitrig} \Omega_{ni} = \int_{(k_t^2+m^2)/s}^1 \frac{d
\alpha}{\alpha} \left(\delta p_t(k_1,k_2)-\delta p_t(k_1+k_2)
\right).
\end{equation}

We shall first address the terms involving $\Omega_0$ the trigger
functions for massive and massless single gluon emission
and then focus on the non-inclusive term.

\section{Dispersive calculations}\label{sec:dispersive}

Here we shall show how the leading non-perturbative terms may be
extracted using the dispersive representation of the running
coupling \cite{DMW}. We start with the inclusive treatment, compute
the relevant trigger functions and then move to the non-inclusive
gluon decay term.

\subsection{Na\"{i}ve result and inclusive correction}

Let us consider the terms in eq.~\eqref{eq:inttrig} involving
$\Omega_0(k_t^2)$ and $\Omega_0(k_t^2+m^2)$. As was discussed in
refs.~\cite{Milan1, Milan2} the integral of the decay matrix element
$M^2$ contains a term proportional to $\beta_0$ as well as a
singular term where the singularity is associated to the collinear
gluon splitting. The collinear singularity cancels against a similar
one in the function $\chi$ leaving a finite remainder. Also the term
proportional to $\beta_0$ combines with the $\alpha_s(0)$ term to
build up the running coupling in the CMW scheme \cite{CMW}, which
shall always be implied henceforth. To be more precise, after
integrating over the decay phase-space $d\Gamma_2$, 
eq.~\eqref{eq:inttrig} can be written as a sum of two finite terms:
\begin{equation}
\left \langle \delta p_t \right \rangle ^{i} = \left \langle\delta
p_t \right \rangle^{0}+\left \langle\delta p_t\right
\rangle^{inc}\,.
\end{equation}

Let us first describe the term $ \left \langle\delta p_t \right
\rangle^{0}$ which reads \cite{Milan2}:
\begin{align}
\label{eq:spl}  \left \langle\delta p_t \right \rangle^{0} &= 4 C_F
\int \frac{dm^2 dk_t^2}{k_t^2+m^2}  \left\{ \frac{\alpha_s(0)}{4
\pi} \delta(m^2) -\frac{\beta_0}{m^2} \left( \frac{\alpha_s}{4 \pi}
\right )^2 \right \} \Omega_0 \left (k_t^2+m^2 \right)\nonumber \\
& = \frac{C_F}{\pi} \int_0^\infty dm^2 \alpha_{{\mathrm{eff}}} (m^2)
\frac{-d}{dm^2} \int_0^{Q^2}\frac{dk_t^2}{k_t^2+m^2} \Omega_0
\left(k_t^2+m^2 \right)  \nonumber\\ & = \frac{C_F}{\pi}
\int_0^\infty \frac{dm^2}{m^2} \alpha_{\mathrm{eff}}(m^2) \Omega_0
(m^2).
\end{align}
In writing the above we have, as in refs.~\cite{DMW, Milan1,
Milan2}, made use of the ultraviolet convergence of the integrals to
extend the upper limits of the $m^2$ integration to infinity, since
we are concerned here with the infrared region $k_t, m \sim
\Lambda_{\mathrm{QCD}}$. We also wrote $Q^2 \approx s$ as the upper
limit of the $k_t$ integral, which is redundant here since our final
results stem from the low $k_t$ infrared regime. The above result
involves invoking the \emph{effective coupling} which is related to
$\alpha_s$ via a dispersive integral \cite{DMW}:
\begin{equation}
\frac{\alpha_s(k^2)}{k^2} = \int_0^{\infty} dm^2
\frac{\alpha_{\mathrm{eff}}(m^2)}{(m^2+k^2)^2}\,,
\end{equation}
so that one has:
\begin{equation}
\label{eq:effco} \frac{d}{d \ln m^2}
\frac{\alpha_{\mathrm{eff}}(m^2)}{4 \pi} = -\beta_0 \left
(\frac{\alpha_s}{4 \pi} \right )^2+\cdots \,, \qquad
\alpha_{\mathrm{eff}}(0) = \alpha_s(0)\,.
\end{equation}
To arrive at the last line of eq.~\eqref{eq:spl} we substituted the
effective coupling instead of $\alpha_s$ using eq.~\eqref{eq:effco}
and performed an integration by parts. The term corresponding to
$\langle \delta p_t \rangle ^0 $ is the one that is produced by
ignoring all two-loop effects except those included via the running
coupling. It thus corresponds to a na\"{i}ve single-gluon treatment
and is thus referred to as the \emph{na\"{i}ve} term. This is the
term that will be related to the one-gluon emission study in
ref.~\cite{Dassalmag}, where a running coupling $\alpha_s(k_t)$ was
introduced by hand into the calculation\footnote{Strictly speaking
the running coupling used in ref.~\cite{Dassalmag} had as argument
the transverse momentum with respect to the dipole axis in the rest
frame of each emitting dipole. For generating the $1/R$ collinear
term it suffices to replace this with $k_t$ the transverse momentum
with respect to the emitting hard leg.}.

The term $\left \langle\delta p_t \right \rangle^{inc}$  is a
leftover from the cancellation of singularities between the function
$\chi$ and the collinear divergent piece of the result for the
integral of $M^2$ \cite{Milan1, Milan2}, which following
refs.~\cite{Milan1,Milan2} we refer to as the inclusive correction
to the na\"{i}ve result:
\begin{align}
\label{eq:inapp} \left \langle\delta p_t\right \rangle^{inc} &=
\frac{8 C_F C_A}{\beta_0} \int \frac{dm^2}{m^2}
\frac{\alpha_{\mathrm{eff}}(m^2)}{4 \pi} \frac{d}{d \ln m^2} \int
\frac{dk_t^2}{k_t^2+m^2} \ln \frac{k_t^2 (k_t^2+m^2)}{m^4}
\Omega^{inc}(k_t^2+m^2), \nonumber \\  \Omega^{inc} &\equiv
\Omega_0(k_t^2+m^2)-\Omega_0(k_t^2).
\end{align}
The inclusive correction merely arises out of the difference between
the massive gluon and massless gluon trigger functions which we
denote by $\Omega^{inc}$, and is the first sign of two-loop effects
not included simply via the use of a running coupling.

\subsubsection{Na\"{i}ve result}

In order to focus on the non-perturbative contribution alone we can
split $\alpha_{s} = \alpha^{\mathrm{PT}}_{s}+ \delta \alpha_{s}$,
with $\delta \alpha_s$ a non-perturbative modification to the
perturbative definition of $\alpha_s$  which results in the
corresponding modification to the effective coupling $\delta
\alpha_{\mathrm{eff}}$.

We can also take the small-$m^2$ limit of the trigger function
$\Omega_0(m^2)$ so that the non-perturbative (NP) contribution
reads:
\begin{equation}
\label{eq:npnaive} \left \langle \delta p_t \right \rangle
^0_{\mathrm{NP}} = \frac{C_F}{\pi} \int_0^\infty \frac{dm^2}{m^2}
\delta \alpha_{\mathrm{eff}}(m^2) \Omega_0 (m^2).
\end{equation}
The non-perturbative modification $\delta \alpha$ must vanish in the
perturbative region so that the coupling $\alpha_{\mathrm{eff}}$
matches on to the perturbative coupling. Hence the integral above,
representing the non-perturbative contribution, has support only
over a limited range of $m^2$ up to some arbitrarily chosen
perturbative matching scale $\mu_I$. The $\mu_I$-dependence of the
result is then cancelled up to ${\mathcal{O}}(\alpha_s^n)$ when
combining the non-perturbative result with a perturbative estimate
at the same order. This issue, mentioned for the sake of
completeness, is irrelevant to the rest of our discussion and shall
henceforth not be mentioned again. We shall also avoid writing the
subscript NP since we always refer to the non-perturbative
contributions from now on.

Let us compute the na\"{i}ve trigger function $\Omega_0(k_t^2+m^2)$
in order to evaluate eq.~\eqref{eq:npnaive}. This is given by:
\begin{equation}
\Omega_0(k_t^2+m^2) = \int_{(k_t^2+m^2)/s}^{1}
\frac{d\alpha}{\alpha} \delta p_t(k),
\end{equation}
where $k=k_1+k_2$ is the parent (massive) gluon four-momentum. Using
the results of the previous section eqs.~\eqref{eq:partin} to
\eqref{eq:partout} we have:
\begin{align}
\delta p_t(k) & = \delta p_t^{+}+\delta p_t^{-} \nonumber\\
&=  - \frac{\sqrt{s}}{{2}}\beta\, \Theta_{\mathrm{in}}(k)
-\frac{\sqrt{s}}{2} \alpha \, \Theta_{\mathrm{out}}(k),
\end{align}
where $ \Theta_{\mathrm{out}}$ and $\Theta_{\mathrm{in}}$ are step
functions indicating whether the gluon is outside or inside the jet
(i.e. if it is recombined with the hard parton or not).

In all the jet algorithms that we consider, the clustering of a
single particle (e.g. a soft gluon) to the triggered hard parton is
based on the distance criterion $\delta \eta^2+\delta \phi^2$, with
$\delta \eta$ and $\delta \phi$ the separation in rapidity and
azimuth between the particle considered and the hard parton. At
small angles (i.e in the collinear limit relevant to our results)
this is just $\theta^2$, where $\theta$ is the small angle between
the soft gluon and the triggered hard parton. Thus at the single
gluon level the soft gluon is combined with the hard parton if
$\theta^2<R^2$, and is not combined if the converse is true.
Moreover using the easily-derived relation\footnote{Strictly
speaking this relation is true for massless gluon clustering. In the
na\"{i}ve calculation we are however free to ignore the gluon mass
in the definition of the clustering angle, since precisely the same
quantity is subtracted in the inclusive and non-inclusive terms at
the end. This freedom to define the role of mass in the na\"{i}ve
calculation was also exploited in refs. \cite{Milan1, Milan2} to
obtain a trigger function $\propto \sqrt{k_t^2+m^2}$.} $\theta^2 = 4
\beta/\alpha$ we have:
\begin{equation}
\Theta_{{\mathrm{out}}}(k) = \Theta \left (4 \beta/\alpha -R^2
\right) = \Theta \left( 2 \frac{\sqrt{k_t^2+m^2}}{R\sqrt{s}} -
\alpha \right),
\end{equation}
where we used $\alpha \beta = (k_t^2+m^2)/s$, and conversely for
$\Theta_{\mathrm{in}}(k)$.

Let us first consider the $\Theta_{\mathrm{out}} = 1$ case or the
gluon not recombined with the jet. The contribution to the trigger
function is:
\begin{equation}
\Omega_0 (k_t^2+m^2) = -\frac{\sqrt{s}}{2}\int_{(k_t^2+m^2)/s}^1
\frac{d \alpha}{\alpha}\, \alpha \,\Theta \left ( 2
\frac{\sqrt{k_t^2+m^2}}{R\sqrt{s}} - \alpha \right).
\end{equation}
Performing the above integral we arrive at:
\begin{equation}
\label{eq:incr} \Omega_0\left (k_t^2+m^2 \right) =
-\frac{\sqrt{k_t^2+m^2}}{R}\,,
\end{equation}
where we retained only the leading ${\mathcal{O}}(k_t/R)$ term we
wish to compute, neglecting terms of order $k_t^2$ that arise from
the lower limit of the integral. Thus the $1/R$ behaviour is
associated to quasi-collinear soft emission such that $1 \gg \alpha
\gg \beta$, where $1 \gg \alpha$ corresponds to soft emission, while
$\alpha \gg \beta$ is the collinear regime. We note that the
convergence of the $\alpha$ integral can be used to extend the
integral to infinity, while the irrelevant lower limit can be set to
zero as in refs.~\cite{Milan1, Milan2}, $\int d\alpha/\alpha \to
\int_0^{\infty} d\alpha/\alpha$. Lastly we just mention that the
term with the soft gluon recombined with the jet, $\delta p_t^{+}$,
does not contribute to the $1/R$ result. The physical reason for
this is straightforward: the contribution to the jet $\delta p_t$
from this term is proportional to $\beta$ or the jet mass, which
vanishes in the collinear limit responsible for the $1/R$ behaviour,
i.e. is collinear-safe. Thus only gluons not recombined to the hard
jet give us a hadronisation correction varying as $1/R$.

Now we substitute eq.~\eqref{eq:incr} into the result
eq.~\eqref{eq:spl} to obtain the na\"{i}ve result:
\begin{equation}
\langle \delta p_t \rangle ^{0} =- \frac{1}{R}\frac{C_F}{\pi}
\int_0^{\infty} \frac{dm^2}{m^2} \delta \alpha_{\mathrm{eff}}(m^2)
m\,.
\end{equation}
This result can be written as
\begin{equation}
\label{eq:finnaive} \langle \delta p_t \rangle ^{0} = -\frac{2
A_1}{R}\,,
\end{equation}
where we introduced the non-perturbative coupling moment $A_1$:
\begin{equation}
A_1 = \frac{C_F}{2 \pi} \int_0^{\infty} \frac{dm^2}{m^2} \,m \,
\delta \alpha_{\mathrm{eff}} (m^2).
\end{equation}

The coefficient $2 A_1$ of the $-1/R$ term can be compared directly
to the na\"{i}ve result for $e^{+}e^{-}$ event-shape variables
\cite{Milan1, Milan2}. It is precisely one-half of the result for
the thrust mean-value exactly as was reported in \cite{Dassalmag}
also within a na\"{i}ve single gluon approach.

\subsubsection{Inclusive correction}

Next we deal with the inclusive correction term. Here the trigger
function is just the difference between the massive and massless
gluon phase-space:
\begin{align}
\Omega^{inc} &= -\frac{\sqrt{s}}{2}\int_0^{\infty} \frac{d
\alpha}{\alpha} \Theta \left (2 \frac{\sqrt{k_t^2+m^2}}{R\sqrt{s}}
-\alpha \right)+\frac{\sqrt{s}}{2} \int_0^{\infty} \frac{d
\alpha}{\alpha} \Theta \left( \frac{2 k_t}{R\sqrt{s}} -\alpha
\right)\nonumber \\  & = -\frac{1}{R} \left(\sqrt{k_t^2+m^2}-k_t
\right).
\end{align}

The inclusive trigger function is precisely of the form one meets in
refs.~\cite{Milan1,Milan2}, with the coefficient of $-1/R$, being
identical to the corresponding quantity for event shape-variables,
enabling us to simply use the results derived for event shapes.
Inserting the trigger function into eq.~\eqref{eq:inapp} and
introducing $\delta \alpha_{\mathrm{eff}}(m^2)$ as before, one gets:
\begin{equation}
\langle \delta p_t \rangle^{inc} = \langle \delta p_t \rangle^{0}
r_{in}\,,\label{eq:rin}
\end{equation}
where $r_{in} =3.299 C_A/\beta_0$ is the value quoted in
refs.~\cite{Milan1, Milan2}. The factor $r_{in}$ is the same as for
the thrust and other event shapes so adding the inclusive correction
to the na\"{i}ve result, the result for the jet $\delta p_t$ is
still one-half of the corresponding result for the thrust, i.e.
universality is maintained at the inclusive level.

Having computed the na\"{i}ve and inclusive pieces we are left with
the non-inclusive correction. It is here that the jet $\delta p_t$
result will no longer receive the same correction as event shapes
and universality of the Milan factor will be broken (except for the
case of the anti-$k_t$ jet algorithm).

\subsection{Non-inclusive correction}

For the non-inclusive term one can follow a similar treatment to the
inclusive case but with the appropriate non-inclusive trigger
function. Then eq.~\eqref{eq:ni} becomes, as for the case of
event-shape variables \cite{Milan1, Milan2},
\begin{equation}
\label{eq:mastni} \langle \delta p_t \rangle^{ni} = \frac{4
C_F}{\beta_0} \int_0^{\infty} \frac{dm^2}{m^2}
\frac{\alpha_{\mathrm{eff}}(m^2)}{4 \pi} \frac{d}{d \ln m^2} \left
\{ \int_0^{2\pi} \frac{d\phi}{2\pi} \int_0^1 dz \int dk_t^2 m^2
\frac{1}{2!} M^2 \Omega_{ni} \right \}.
\end{equation}
We shall focus our attention on the term in parenthesis that yields
the linear behaviour in $m$ corresponding to the leading
non-perturbative (hadronisation) correction. In order to compute
this we need to evaluate the non-inclusive trigger function:
\begin{equation}
\int_0^{\infty} \frac{d\alpha}{\alpha} \left( \delta p_t(k_1,k_2)-
\delta p_t(k_1+k_2) \right).
\end{equation}
The term $\delta p_t(k_1,k_2)$ represents the contribution from two
parton emission, which depends on the details of whether the
offspring partons end up inside or outside the jet \emph{after} the
application of the jet algorithm. For most jet algorithms currently
in use, this involves more than just working out whether the gluons
are within a distance $R$ of the hard parton (an exception is the
anti-$k_t$ algorithm), so we shall need to consider the action of
the algorithms in some detail. Due to the length of this discussion
we shall devote the next section entirely to it.

\section{The non-inclusive trigger function}\label{sec:ni-trigger}

We shall start by considering the $k_t$ algorithm. Let us briefly
remind the reader of how this works.

First one computes the distances $d_{ij}$ between all pairs of
objects $i$ and $j$, $d_{ij}
=\mathrm{min}\left(\kappa_{t,i}^2,\kappa_{t,j}^2 \right)y_{ij}$,
with $y_{ij} = \delta \eta^2 +\delta \phi^2$, where $\delta \eta$
and $\delta \phi$ refer to the separation of $i$ and $j$ in rapidity
and azimuth measured with respect to the beam and we used $\kappa_t$
to denote the transverse momentum with respect to the beam direction
(while reserving $k_t$ for the transverse momentum with respect to
the hard emitting parton). In the small angle approximation
appropriate to our work we just have $\delta \eta^2+\delta\phi^2 \to
\theta_{ij}^2$, where $\theta_{ij}$ is the angle between $i$ and
$j$. One also computes the distances $d_{iB}$ between each object
and the beam defined as $\kappa_{t,i}^2 R^2$, with $R$ the selected
radius parameter.

Then if amongst the various distances the smallest is a $d_{iB}$,
object $i$ is a jet and is removed from the list of objects to be
clustered. If the smallest distance is a $d_{ij}$ then $i$ and $j$
are combined (merged) into a single object and the procedure is then
iterated until all objects have been removed. The recombination
scheme we use here is addition of four-momenta so that the object
resulting from combining $i$ and $j$ has four-momentum $p_i+p_j$.

We note from eqs.~\eqref{eq:partin} to \eqref{eq:partout} that the
$\delta p_t(k_1,k_2)$ is additive over the contributions of $k_1$
and $k_2$,
\begin{align}
\delta p_t (k_1,k_2) & = \delta p_t(k_1)+\delta p_t(k_2)\nonumber\\
&= \delta p_t^+(k_1) \Xi_{\mathrm{in}}(k_1)+\delta p_t^-(k_1)
\Xi_{\mathrm{out}}(k_1)+\left (k_1 \leftrightarrow k_2 \right),
\end{align}
where we expressed $\delta p_t$ in terms of the contributions from
$k_1$ and $k_2$. We also distinguished the cases when the offspring
partons are recombined into and fly outside the jets as the
contributions are different in either case:
\begin{align}
\delta p_t^{+}(k_i) & = -\frac{\sqrt{s}}{{2}} \beta_i\,,\nonumber \\
\delta p_t^{-}(k_i) &= -\frac{\sqrt{s}}{{2}} \alpha_i\,,
\end{align}
where $p_t^+$ expresses the contribution from a recombined parton
and $p_t^-$ for a non-recombined emission. The conditions
$\Xi_{\mathrm{in,out}}$ denote the constraints for partons to be in
and outside the jets and they are no longer simple step functions as
for the single gluon case, and must be obtained by applying the
algorithm in full including the possibility of soft partons
clustering to each other.

\subsection{$\delta p_t$ in the $k_t$ algorithm}

Let us first apply the $k_t$ algorithm to work out the $\delta
p_t(k_1,k_2)$ in various situations. We have to consider the soft
partons $k_1$, $k_2$ and their distances from the triggered hard
parton ``jet'' which we denote by $j$. To this end let us define the
step functions $\Theta_{ab} = \Theta \left(R^2-\theta_{ab}^2
\right)$, so that if two partons have an angular separation less
than $R$ the step function is unity and else it is
zero\footnote{Recall that we are looking at the small-$R$
behaviour $R \ll 1$, which entitles us to replace $\delta
\eta^2+\delta \phi^2$ by the small angle approximation $\theta^2$.
Neglected terms produce only finite $R$ corrections.}. Then we can
divide the full phase-space up into definite regions (as indicated
by table \ref{tab:clustering}) which account for all possible cases
for the angular separations. For each row of the table (i.e each of
the possible combinations of values of the $\Theta_{ab}$, it is
possible to obtain a value for the $\delta p_t(k_1,k_2)$.
\TABLE[ht]{\parbox{0.6\textwidth}{
\begin{center}
\begin{tabular}{|c|c|c|c|}
\hline $\Theta_{12}$ & $\Theta_{j1}$ & $\Theta_{j2}$ &
$\delta p_t(k_1,k_2)$\\
\hline
1 & 1 & 1 &  $\delta p_t^{++}(k_1,k_2)$\\
\hline
1 & 1 & 0 & Discussed below\\
\hline
1 & 0 & 1 & Discussed below\\
\hline
1 & 0 & 0 & Discussed below\\
\hline
0 & 1 & 1 & $\delta p_t^{++}(k_1,k_2)$\\
\hline
0 & 1 & 0 & $\delta p_t^{+-}(k_1,k_2)$\\
\hline
0 & 0 & 1 & $\delta p_t^{-+}(k_1,k_2)$\\
\hline
0 & 0 & 0 &  $\delta p_t^{--}(k_1,k_2)$\\
\hline
\end{tabular}
\caption{\label{tab:clustering} Values of $\delta p_t$ depending on
the angular separations of partons. The three entries ``Discussed
below'' need special treatment and thus are left for discussion in
the text.}
\end{center}
}}

In table \ref{tab:clustering} we have written $\delta
p_t^{++}(k_1,k_2) = \delta p_t^+(k_1)+\delta p_t^+(k_2)$ to indicate
the contribution when both $k_1$ and $k_2$ are eventually combined
into the jet, while $\delta p_t^{+-}(k_1,k_2) = \delta
p_t^+(k_1)+\delta p_t^-(k_2)$ indicates a situation where $k_1$ is
recombined and $k_2$ is not. Let us show how the entries in this
table are obtained by considering an example and discussing the more
interesting scenarios.

\begin{description}
\item[The scenario: $\Theta_{12}=1$, $\Theta_{j1}=1$,
$\Theta_{j2}=0$.] We consider the case described in the second row
of the above table. The distances we need to discuss are the
following:
\begin{eqnarray}
d_{1B}&=&\kappa_{t,1}^2 R^2\,, \nonumber\\
d_{2B}&=&\kappa_{t,2}^2 R^2\,,\nonumber\\
d_{12}&=&\min(\kappa_{t,1}^{2},\kappa_{t,2}^2)\theta_{12}^2\,,
\nonumber\\
d_{1j}&=&\kappa_{t,1}^2 \theta_{1j}^2\,,\nonumber\\
d_{2j}&=&\kappa_{t,2}^2 \theta_{2j}^2\,.\nonumber
\end{eqnarray}

In this situation (given the values of the $\Theta$ functions) it is
not possible for any of $d_{1B}$, $d_{2B}$, or $d_{2j}$ to be the
smallest. Thus either $d_{12}$ or $d_{1j}$ is the smallest.

If $d_{12}$ is the smallest (i.e. smaller than $d_{1j}$) then $k_1$
and $k_2$ are clustered to each other first. Two sub-scenarios
arise: the resultant soft jet may fall either within or outside the
grasp of the hard parton. If the soft jet is recombined with the
hard parton the resulting contribution is $\delta p_t^{++}$, 
else it is $\delta p_t^{--}$.

Alternatively  if $d_{1j}$ is the smallest distance particle $k_1$
is combined with the jet first. Then particle $k_2$ forms a separate
jet, i.e. is removed as it is closer to the beam than it is to the
hard jet. The contribution is then $\delta p_t^{+-}(k_1,k_2)$. We
can summarise the overall scenario just discussed by the result:
\begin{description}
\item[Case $d_{12}<d_{1j}$:]
\begin{equation}
\delta p_t(k_1,k_2) = \Theta(d_{1j}-d_{12})\left[
\Theta(R^2-\theta^2_{kj}) \delta p_t^{++}
+\Theta(-R^2+\theta^2_{kj}) \delta p_t^{--}\right],
\end{equation}
where $\theta_{kj}$ is the angle of the soft jet (parent gluon)
$k=k_1+k_2$ with the hard parton.
\item[Case $d_{12}>d_{1j}$:] In this case particle 1 gets clustered
to the jet and 2 is left outside. Thus:
\begin{equation}
\delta p_t(k_1,k_2) = \Theta(d_{12}-d_{1j}) \delta
p_t^{+-}(k_1,k_2).
\end{equation}
\end{description}

\item[The scenario: $\Theta_{12}=1$, $\Theta_{j1}=0$,
$\Theta_{j2}=1$.] This scenario (third row of the table above) is
the same as the one just discussed but with $k_1$ and $k_2$
interchanged.

\item[The scenario: $\Theta_{12}=1$, $\Theta_{j1}=0$
, $\Theta_{j2}=0$.] Here the smallest quantity is $d_{12}$ and the
soft partons are clustered to each other first. The resulting soft
jet may or may not be recombined with the hard parton. Thus we have
in this region:
\begin{equation}
\delta p_t(k_1,k_2) = \Theta(R^2-\theta^2_{kj}) \delta p_t^{++}
+\Theta(-R^2+\theta^2_{kj}) \delta p_t^{--}\,,
\end{equation}
where the step functions specify whether the soft parton jet is
within an angular distance $R$ of the  hard parton or not.
\end{description}

All the other scenarios follow straightforwardly and the values of
the $\delta p_t$ are as indicated in the table. For instance if one
considers the bottom row of the table where all the $\Theta$
functions vanish (partons always separated by more than $R$) the
$d_{iB}$ are always smallest, and the partons $k_1$ and $k_2$ are
never combined in to the triggered jet, so the result is $\delta
p_t^{--}$.

Now we are in position to write down the expression for $\delta p_t$
at two-parton level. We express it as follows:
\begin{eqnarray}\label{eq:det}
\delta p_t (k_1,k_2) &= & \delta p_t^{++} (k_1,k_2)
\Theta_{\mathrm{in}}(k_1) \Theta_{\mathrm{in}}(k_2)+\nonumber\\
&+&\delta p_t^{--} (k_1,k_2)\Theta_{\mathrm{out}}(k_1)
\Theta_{\mathrm{out}}(k_2)\left[1-\Theta_{12}(k_1,k_2)+
\Theta_{12}(k_1,k_2)\Theta_{\mathrm{out}}(k)\right]+\nonumber\\
&+&\delta p_t^{++}(k_1,k_2) \Theta_{\mathrm{out}}(k_1)
\Theta_{\mathrm{out}}(k_2) \Theta_{12}(k_1,k_2)
\Theta_{\mathrm{in}}(k)+\nonumber\\
&+&\delta p_t^{++}(k_1,k_2) \Theta_{\mathrm{in}}(k_1)
\Theta_{\mathrm{out}} (k_2) \Theta_{12}(k_1,k_2)
\Theta(d_{1j}-d_{12})\Theta_{\mathrm{in}}(k)
+(k_1\leftrightarrow k_2)+\nonumber\\
&+&\delta p_t^{--}(k_1,k_2) \Theta_{\mathrm{in}}(k_1)
\Theta_{\mathrm{out}} (k_2) \Theta_{12}(k_1,k_2)
\Theta(d_{1j}-d_{12}) \Theta_{\mathrm{out}}(k)
+(k_1\leftrightarrow k_2)+\nonumber\\
&+&\delta p_t^{+-}(k_1,k_2) \Theta_{\mathrm{in}}(k_1)
\Theta_{\mathrm{out}} (k_2) \Theta_{12}(k_1,k_2)
\Theta(d_{12}-d_{1j})+(k_1\leftrightarrow k_2)+\nonumber\\
&+&\delta p_t^{+-}(k_1,k_2) \Theta_{\mathrm{in}}(k_1)
\Theta_{\mathrm{out}}(k_2) (1-\Theta_{12}) +(k_1 \leftrightarrow
k_2).
\end{eqnarray}
In the above $\Theta_{\mathrm{in}}(k_i) = \Theta(R^2-\theta_{ij}^2)$
and $\Theta_{\mathrm{out}} = 1-\Theta_{\mathrm{in}}$ pertains to
whether the parton $k_i$ is within or outside an angular separation
$R$ of the hard parton, while $\Theta_{12}= \Theta(R^2- \theta_{1
2}^2)$ is the constraint for the soft partons to have an angular
separation less than $R$. Eq.~\eqref{eq:det} merely summarises the
contents of the table, for instance the sum of entries in the first
and fifth rows is just the first line of the above equation. For
consistency one can check by removing all the $\delta p_t$ terms in
the above equation and adding all the theta functions as they appear
one should get 1. The check is positive (meaning all possible
scenarios have been discussed without double counting).

We can also write eq.~\eqref{eq:det} in terms of the individual
contributions of $k_1$ and $k_2$ since the $\delta p_t$ is additive
over the contributions of individual partons:
\begin{equation}
\delta p_t(k_1,k_2) = \delta p_t(k_1) + \delta p_t(k_2) =  \delta
p_t^{-}(k_1) \Xi_{\mathrm{out}}(k_1)+ \delta p_t^{+}(k_1)
\Xi_{\mathrm{in}}(k_1)+(k_1 \leftrightarrow k_2),
\end{equation}
where we separated the contributions from each offspring parton into
its $+$ and $-$ components according to whether the parton is
recombined with the jet or not.

Next we note that for the $1/R$ term one is only interested in
partons not recombined to the jet. Partons recombined with the jet
contribute to the $\delta p_t$ via $\delta p_t^+(k_i) \propto
\beta_i$. Integrating such contributions over the $d\alpha/\alpha$
phase-space produces only terms regular in $R$, as for the single
gluon case. Thus for computing the $1/R$ term the $\delta
p_t(k_1,k_2)$ can just be expressed as:
\begin{equation}
\delta p_t(k_1,k_2) = \delta p_t^{-}(k_1) \Xi_{\mathrm{out}}(k_1)+
\delta p_t^{-}(k_2) \Xi_{\mathrm{out}}(k_2).
\end{equation}

Thus from eq.~\eqref{eq:det} focussing on all terms where $k_1$ is
not in the jet we can identify:
\begin{eqnarray}
\Xi_{\mathrm{out}} (k_1) &= & \Theta_{\mathrm{out}}(k_1)
\Theta_{\mathrm{out}}(k_2) \left[1-\Theta_{12}(k_1,k_2)
+\Theta_{12}(k_1,k_2)\Theta_{\mathrm{out}}(k) \right]+\nonumber\\
&+& \Theta_{\mathrm{in}}(k_1) \Theta_{\mathrm{out}} (k_2)
\Theta_{12}(k_1,k_2) \Theta(d_{1j}-d_{12})
\Theta_{\mathrm{out}}(k)+\nonumber\\
&+& \Theta_{\mathrm{in}}(k_2)\Theta_{\mathrm{out}} (k_1)
\Theta_{12}(k_1,k_2) \Theta(d_{2j}-d_{12})
\Theta_{\mathrm{out}}(k)+\nonumber\\
&+& \Theta_{\mathrm{in}}(k_2)\Theta_{\mathrm{out}} (k_1)
\Theta_{12}(k_1,k_2) \Theta(d_{12}-d_{2j})+\nonumber\\
&+& \Theta_{\mathrm{in}}(k_2) \Theta_{\mathrm{out}}(k_1)
(1-\Theta_{12}),
\end{eqnarray}
which can be simplified after some algebra to:
\begin{eqnarray}
\Xi_{\mathrm{out}} (k_1) &= & \Theta_{\mathrm{out}} (k_1)\left[1-
\Theta_{\mathrm{out}}(k_2)\Theta_{12}(k_1,k_2)
\Theta_{\mathrm{in}}(k)\right]+\nonumber\\
&+& \Theta_{\mathrm{in}}(k_1)\Theta_{\mathrm{out}} (k_2)
\Theta_{12}(k_1,k_2) \Theta(d_{1j}-d_{12})
\Theta_{\mathrm{out}}(k)+\nonumber\\
&-& \Theta_{\mathrm{out}} (k_1)\Theta_{\mathrm{in}}(k_2)
\Theta_{12}(k_1,k_2)\Theta(d_{2j}-d_{12})\Theta_{\mathrm{in}}(k),
\end{eqnarray}
which represents all possible ways for $k_1$ to be outside the jet
after application of the $k_t$ algorithm. A similar expression holds
for $k_2$.

Repeating the above procedure for the anti-$k_t$ algorithm yields a
much simpler result due to the absence of soft parton
self-clustering. Thus in the anti-$k_t$ case a given parton $k_i$ is
always outside the jet if it is separated by more than $R$ in angle
from the hard parton:
\begin{equation}
\Xi_{\mathrm{out}} (k_i) = \Theta_{\mathrm{out}}(k_i).
\end{equation}

\subsection{The $\alpha$ integral}

Having worked out the relevant piece of the $\delta p_t(k_1,k_2)$
for the $k_t$ algorithm we now proceed to carry out the integral
over $\alpha$ involved in the definition of the trigger function
eq.~\eqref{eq:nitrig}. This is given by:
\begin{eqnarray}
\Omega_{ni} &=& \int \frac{d\alpha}{\alpha} \left( \delta p_t^-(k_1)
\Xi_{\mathrm{out}}(k_1)+ \delta p_t^- (k_2)\Xi_{\mathrm{out}}(k_2)
-\delta p_t^-
(k) \Theta_{\mathrm{out}} (k)\right)\nonumber\\
&=& -\frac{\sqrt{s}}{2} \int \frac{d\alpha}{\alpha} \left(\alpha_1
\Xi_{\mathrm{out}}(k_1)+\alpha_2 \Xi_{\mathrm{out}}(k_2)-\alpha
\Theta_{\mathrm{out}}(k)\right)\nonumber\\
&=&-\frac{\sqrt{s}}{2}\int d\alpha \left(z
\Xi_{\mathrm{out}}(k_1)+(1-z)\Xi_{\mathrm{out}}(k_2)-
\Theta_{\mathrm{out}}(k)\right)\nonumber\\
&=& \Omega_1+\Omega_2-\Omega_3 \,,
\end{eqnarray}
where
\begin{eqnarray}
\Omega_1 &=& -\frac{\sqrt{s}}{2} z \int_0^\infty d\alpha
\Big\{\Theta_{\mathrm{out}} (k_1)\left[1- \Theta_{\mathrm{out}}
(k_2)\Theta_{12}(k_1,k_2)\Theta_{\mathrm{in}}(k)\right]+\nonumber\\
&&\qquad\qquad\qquad+ \Theta_{\mathrm{in}}(k_1)\Theta_{\mathrm{out}}
(k_2)\Theta_{12}(k_1,k_2)\Theta(d_{1j}-d_{12})
\Theta_{\mathrm{out}}(k) +\nonumber\\
&&\qquad\qquad\qquad- \Theta_{\mathrm{out}} (k_1)
\Theta_{\mathrm{in}} (k_2) \Theta_{12}(k_1,k_2)
\Theta(d_{2j}-d_{12})\Theta_{\mathrm{in}}(k)\Big\},
\end{eqnarray}
with eg. $\Theta_{\mathrm{out}}(k_1) = \Theta(\theta_{1j}^2-R^2)$.
Note that in writing the above we have used $\delta p_t^-(k_i)
=-\alpha_i \sqrt{s}/2 $, with $\alpha_1 = z \alpha$ and $\alpha_2
=(1-z) \alpha$. We shall give expressions for $\Omega_2$ and
$\Omega_3$ in due course.

In order to proceed we shall also need to express the angles between
partons in terms of the Sudakov variables. The angle between the
soft parton $k_1$ and the hard parton $j$ is just (assuming
$\alpha_i \gg \beta_i$, which is the collinear approximation)
$\theta_{1j}^2 = 4 \beta_1/\alpha_1 = 4 k_{t,1}^2/(s \alpha_1^2)$.
Moreover since $\alpha_1 = z \alpha$, $\theta_{1j}^2 = 4 q_{1}^2/(s
\alpha^2)$, where $q_1 = k_{t1}/z$. Similarly we have $\theta_{2j}^2
= 4 q_{2}^2/(s \alpha^2)$ with $q_2=k_{t2}/(1-z)$. The angle between
the soft jet formed by clustering $k_1$ and $k_2$ and the hard
parton is given (again in the collinear limit) by $\theta_{kj}^2 = 4
k_t^2/(s \alpha^2)$, and finally the angle between $k_1$ and $k_2$
is just $\theta_{12}^2 = 4 q^2/(s \alpha^2)$, where $q^2 =
(\vec{k}_{t1}/z-\vec{k}_{t2}/(1-z) )^2 $. In writing these angles we
have used the small angle approximation throughout, specifically
$(1-\cos\theta)\approx \theta^2/2$, and also considered $\alpha_i
\gg \beta_i$ as we are always concerned with partons in the region
collinear to the emitting hard parton. Terms neglected in this
approximation contribute to corrections regular in $R$.

Next we re-scale $\sqrt{s} R \alpha/2\rightarrow \alpha $ so as to
straightforwardly extract the $1/R$ dependence that concerns us.
Hence we write:
\begin{eqnarray}
\label{eq:triginvr}\Omega_1&=& -\frac{1}{R} z \int_0^\infty d\alpha
\Big\{ \Theta\left(q_1-\alpha\right)\left[1- \Theta
\left(q_2-\alpha\right)\Theta\left(\alpha-q\right)
\Theta\left(\alpha-k_{t}\right)\right]+\nonumber\\
&&\!\!\qquad\qquad+\Theta\left(\alpha-q_1\right)
\Theta\left(q_2-\alpha\right) \Theta\left(\alpha-q\right)
\Theta\left(k_{t}-\alpha\right) \Theta\left(\kappa_{t,1}^2q_1^2
-\mathrm{min}\left[\kappa_{t,1}^2,
\kappa_{t,2}^2\right] q^2\right)+\nonumber\\
&&\!\!\qquad\qquad- \Theta\left(q_1-\alpha\right)
\Theta\left(\alpha-q_2\right) \Theta\left(\alpha- q\right)
\Theta\left(\alpha-k_{t}\right) \Theta
\left(\kappa_{t,2}^2q_2^2-\mathrm{min} \left[\kappa_{t,1}^2,
\kappa_{t,2}^2\right]q^2\right)\Big\}\nonumber\\
&=& -\frac{1}{R}z\Big\{q_1-\left(\mathrm{min}\left[q_1,q_2\right]
-\mathrm{max}\left[q, k_{t}\right]\right) \Theta\left(
\mathrm{min}\left[q_1,q_2\right] -\mathrm{max}\left[q,
k_{t}\right] \right)+\nonumber\\
&+&\left(\mathrm{min}\left[q_2,k_{t}\right] -\mathrm{max}
\left[q_1,q\right] \right) \Theta\left( \mathrm{min}
\left[q_2,k_{t}\right] -\mathrm{max}\left[ q_1,q\right]
\right)\nonumber\Theta \left(q_1^2 -q^2 \mathrm{min}\left[1,
\kappa_{t,2}^2/ \kappa_{t,1}^2\right] \right) +\nonumber\\&-&
\left(q_1 -\mathrm{max} \left[q_2, q,k_{t} \right] \right)\Theta
\left(q_1- \mathrm{max}\left[q_2, q,k_{t} \right]
\right)\Theta\left(q_2^2-q^2 \mathrm{min} \left[
\kappa_{t,1}^2/\kappa_{t,2}^2,1\right]\right) \Big\},
\end{eqnarray}
where we used the fact that the distances $d_{ij}$ between partons
involve the transverse momenta $\kappa$ with respect to the beam
direction which must be distinguished from $k_t$, the transverse
momentum with respect to the jet direction, and we performed the
trivial $\alpha$ integral extracting the crucial $1/R$ dependence we
seek.

Note that the transverse momenta with respect to the beam are
essentially the longitudinal momenta with respect to the jet (recall
that we are looking at jet production at ninety degrees to the beam
as well as parton emission very close to the triggered hard parton,
$\alpha_i \gg \beta_i$), so we can to our accuracy, 
just use $\kappa_{t,1} \approx
\sqrt{s}/2\, \alpha_1 = \sqrt{s}/2\,z \alpha$ and likewise for
$\kappa_{t,2}$ with $z \to 1-z$.

Thus we can express eq.~\eqref{eq:triginvr} as:
\begin{eqnarray}
\Omega_1&=& -\frac{1}{R} z \Big\{q_1- \Psi\left( \mathrm{min} \left[
q_1,q_2\right]-\mathrm{max}\left[q,k_{t} \right]\right)\nonumber+\\
&&\!\quad\qquad+\Psi\left( \mathrm{min}\left[q_2, k_{t}\right]
-\mathrm{max} \left[q_1,q\right] \right) \times \Theta \left(
q_1-q\,\mathrm{min}\left[1,(1-z)/z\right]\right)+\nonumber\\
&&\!\quad\qquad- \Psi\left(q_1-\mathrm{max}\left[q_2,
q,k_{t}\right]\right) \times \Theta\left(q_2-q\,\mathrm{min}
\left[z/(1-z),1\right]\right)\Big\},
\end{eqnarray}
where we defined the function $\Psi(x) = x\,\Theta(x)$. Here
$\Psi(x)=x$ if $x>0$ and $\Psi(x)=0$ otherwise.

A similar result is obtained for the $\delta p_t(k_2)$ contribution:
\begin{eqnarray}
\Omega_2&=& -\frac{1}{R} (1-z) \Big\{q_2- \Psi\left( \mathrm{min}
\left[
q_1,q_2\right]-\mathrm{max}\left[q,k_{t} \right]\right)\nonumber+\\
&&\!\quad\qquad+\Psi\left( \mathrm{min}\left[q_1, k_{t}\right]
-\mathrm{max} \left[q_2,q\right] \right) \times \Theta \left(
q_2-q\,\mathrm{min}\left[1,z/(1-z)\right]\right)+\nonumber\\
&&\!\quad\qquad- \Psi\left(q_2-\mathrm{max}\left[q_1,
q,k_{t}\right]\right) \times \Theta\left(q_1-q\,\mathrm{min}
\left[(1-z)/z,1\right]\right)\Big\}.
\end{eqnarray}

The subtracted $\Omega_{3}$ contribution is simply the subtraction
of the na\"{i}ve trigger function computed before:
\begin{eqnarray}
\Omega_3 &=& -\frac{1}{R } \sqrt{k_{t}^2+m^2}.
\end{eqnarray}

The overall result for the non-inclusive trigger function is then
given by combining the $\Omega_1$, $\Omega_2$ and $\Omega_{3}$
terms and takes the form:
\begin{equation}
\Omega_{ni} = -\frac{1}{R} \left(k_{t1}+k_{t2}-\sqrt{k_{t}^2+m^2} +
f(q_1,q_2,z)\right),
\end{equation}
which we wrote in terms of the $k_{ti}$ for sake of easier
comparison with the corresponding expression eq.~(3.31) in
ref.~\cite{Milan2}. The first three terms of the result correspond
to the non-inclusive trigger function computed for event shapes
while the extra term $f(q_1,q_2,z)$ will break the universality of
the Milan factor.

To express the result in terms familiar from
refs.~\cite{Milan1,Milan2} we introduce $u_1$ and $u_2$, where
$u_i=q_i/q$. Defining $\widetilde \Omega (u_1,u_2)$ so that
$\Omega_{ni} (q_1,q_2) = -q/R\, \widetilde{\Omega}(u_1,u_2)$ we
arrive at:
\begin{eqnarray}
\widetilde{\Omega}(u_1,u_2)&=&zu_1+(1-z)u_2-\sqrt{zu_1^2+(1-z)u_2^2}
+\nonumber\\
&-&\Psi\left(\mathrm{min}[u_1,u_2]-\mathrm{max}\left[1,
\sqrt{zu_1^2+(1-z)u_2^2-z(1-z)}\right]\right)+\nonumber\\
&+&z \bigg\{\Psi\left( \mathrm{min}\left[u_2, \sqrt{zu_1^2+
(1-z)u_2^2-z (1-z)}\right]-\mathrm{max}\left[u_1,1\right]\right)
\times \nonumber\\
&&\qquad\times\Theta\left(u_1-\mathrm{min}
\left[1,(1-z)/z\right]\right)+\nonumber\\
&&\qquad- \Psi\left(u_1-\mathrm{max}\left[u_2, 1,\sqrt{zu_1^2+(1-z)
u_2^2 -z(1-z)} \right]\right)\times \nonumber \\
&& \qquad\times \Theta\left(u_2-\mathrm{min}\left[z/(1-z),1
\right]\right)\bigg\}+\nonumber\\
&+&(1-z)\bigg\{\Psi\left(\mathrm{min}\left[u_1,\sqrt{zu_1^2+(1-z)
u_2^2-z(1-z)}\right]-\mathrm{max}\left[u_2,1\right]
\right)\times \nonumber\\
&&\qquad\qquad\times \Theta\left(u_2-\mathrm{min}
\left[z/(1-z),1\right]\right)+\nonumber\\
&&\qquad\qquad- \Psi\left(u_2-\mathrm{max}\left[u_1,
1,\sqrt{zu_1^2+(1-z)u_2^2-z(1-z)}\right]\right) \times \nonumber\\
&& \qquad\qquad\times \Theta\left(u_1-\mathrm{min}
\left[1,(1-z)/z\right]\right)\bigg\},\label{eq:trigger}
\end{eqnarray}
where we used $(k_t^2+m^2 )/q^2 = zu_1^2+(1-z)u_2^2$, which follows
from the fact that $m^2 =z(1-z)q^2$.

Once again the first line of the above result corresponds to the
Milan factor for event-shape variables, and is the same trigger
function as encountered in refs.~\cite{Milan1, Milan2}, while the
remainder will constitute the correction to it from $k_t$
clustering.

The result corresponding to the anti-$k_t$ algorithm is also just
the first line of the above result \eqref{eq:trigger} since there we
have:
\begin{equation}
\label{eq:antkt} \Omega_1 = -\frac{1}{R} z \int \frac{d
\alpha}{\alpha} \, \alpha \, \Theta \left (q_1-\alpha \right),
\end{equation}
which gives $-q/R \,(z u_1)$. Similarly the $\Omega_2$ term is the
same with $z \to 1-z$ and the na\"{i}ve piece $\Omega_3$ is of
course common to both algorithms.

Thus in the anti-$k_t$ algorithm after combining all the pieces we
obtain that the result will be identical to that for event-shape
variables, computed in refs. \cite{Milan1, Milan2}.

Having derived the trigger function in the $k_t$ algorithm we now
proceed to perform the integration over the remaining variables,
including also the decay matrix element $M^2$ as required by
eq.~\eqref{eq:mastni}. The details of the integration procedure will
be consigned to the appendix and in the next section we shall quote
the result and combine it with the na\"{i}ve and inclusive terms.

\section{Results and conclusions}\label{sec:conc}

In this section we will mention the result obtained for the
non-inclusive correction using the trigger function of the previous
section. We shall then combine the result with the na\"{i}ve and
inclusive pieces computed before.

After computing the trigger function $\Omega_{ni}$ described in the
previous section one inserts the result into eq.~\eqref{eq:mastni}
to obtain the result for $\langle \delta p_t \rangle^{ni}$. In order
to do so we need to integrate over the parton decay phase-space
including the squared matrix element $M^2$ describing gluon decay.
The details of this integration are mentioned in the appendix and
here we just quote the result analogous to eq.~(4.8) of
\cite{Milan2}:
\begin{equation}
\langle \delta p_t \rangle ^{ni} = \langle \delta p_t \rangle ^0
r_{ni}\,,
\end{equation}
where for the $k_t$ algorithm we have a result for $r_{ni}$
different from that for the event-shape variables studied in
refs.~\cite{Milan1, Milan2}:
\begin{eqnarray}
r_{ni} & = & \frac{2}{\beta_0}
(-2.145C_A+0.610C_A-0.097n_f)\nonumber\\
&=& \frac{1}{\beta_0} (-3.071C_A-0.194n_f),
\end{eqnarray}
where the results in the first line correspond to the ``soft
gluon'', ``hard gluon'' and ``hard quark'' contributions in the
terminology of refs.~\cite{Milan1,Milan2}.

The corresponding result for the anti-$k_t$ algorithm coincides with
that obtained for event shapes~\cite{Milan1, Milan2}:
\begin{equation}
r_{ni} =  \frac{2}{\beta_0} (-1.227 C_A+0.365C_A-0.052 n_f).
\end{equation}

Combining the result obtained for $r_{ni}$ with that for the
inclusive piece (eq.~\eqref{eq:rin}) and the na\"{i}ve result (eq.
\eqref{eq:finnaive}) one obtains the full result for jets defined in
the $k_t$ algorithm:
\begin{eqnarray}
\langle \delta p_t \rangle & =& \frac{-2 A_1}{R}
\left(1+r_{in}+r_{ni}\right)
\nonumber \\
&=& \frac{-2 A_1}{R} (1.01),
\end{eqnarray}
for $n_f =3$ (for $n_f=0$ the rescaling factor is $1.06$). The
factor multiplying the na\"{i}ve result is thus found to be $1.01$
compared to $1.49$ for event shapes \cite{Milan1, Milan2} and the
anti-$k_t$ algorithm. Here we took $n_f=3$ since we are dealing with
the non-perturbative region but in the corresponding perturbative
estimates one should of course take $n_f =5$. To obtain the results
for a gluon jet one can just replace $C_F$ in $A_1$, by $C_A$.

Thus compared to the thrust $1/Q$ correction which has a Milan
factor of $1.49$ the $1/R$ term of the jet hadronisation for quark
jets in the $k_t$ algorithm has a coefficient that is $0.5 \times
1.01/1.49 \approx 0.34$ times the coefficient of $1/Q$ for
$e^{+}e^{-}$ thrust instead of $0.5$ as observed at the single gluon
level in \cite{Dassalmag}. For the anti-$k_t$ algorithm the result
is still one-half of that for the thrust after inclusion of the
Milan factor $1.49$ in both cases. The somewhat smaller
hadronisation correction predicted here for the $k_t$ algorithm
compared to the anti-$k_t$ algorithm was also observed in the
Monte-Carlo studies of Ref.~\cite{Dassalmag}.

To summarise, we began this paper by noting that the single-gluon
calculations performed in ref.~\cite{Dassalmag} are inadequate in
terms of determining the size of the hadronisation correction to jet
$p_t$ relative to the known $1/Q$ corrections for event shapes. The
reason for this inadequacy was merely the fact that a two-loop
calculation had been shown to be necessary in
refs.~\cite{Milan1,Milan2} to unambiguously determine the size of
$1/Q$ hadronisation corrections to event shapes.

Our aim in this paper was to carry out a similar calculation for the
jet $p_t$ so as to finally put it on the same footing as event shape
variables, which makes the comparison feasible. We carried out such
a calculation and pointed out that as for event-shape variables one
can write the result as a two-loop rescaling factor multiplying the
na\"{i}ve single-gluon result. \emph{Unlike} event-shape variables
however, for jets defined in the $k_t$ algorithm the rescaling
factor is not the universal Milan factor $1.49$ turning out instead
to be $1.01$. We confirmed the expectation of ref.~\cite{antikt}
that for jets defined in the anti-$k_t$ algorithm the rescaling
factor is $1.49$, as for event-shape variables.

We also note that we have presented here the calculation of
correlated two-parton emission and neglected higher correlations
involving three or more partons which would come in beyond our
two-loop accuracy. However the two-loop calculation we performed
here corresponds to the accuracy needed to fix the scale of the
coupling or equivalently to fix a value for $\Lambda_{\mathrm{QCD}}$
which directly controls the size of the non-perturbative
contributions. Hence while one expects the effects of the two-loop
calculation to be crucial, higher-loop calculations can be expected
to contribute at most corrections of relative order $\alpha_s/\pi$
which could change the value of the two-loop factors at most at the
twenty percent level \cite{Milan1}. Interestingly Sterman and Lee
\cite{slee} have recently shown within the context of soft-collinear
effective theory that for observables such as event-shape variables
that are linear in emitted parton transverse momenta, one can
generally demonstrate universality of power corrections without
resorting to fixed-order perturbative calculations beyond the
na\"{i}ve massive gluon results. It would be of interest to revisit
their arguments in the context of the observable studied in the
current paper.

To conclude we point out that the potential experimental studies
mentioned in \cite{Dassalmag} should of course take into account the
result of the calculation performed here for jets defined in the
$k_t$ and anti-$k_t$ algorithms. The calculation for jets defined in
the SISCone and the Cambridge/Aachen algorithms is work in progress
\cite{dasdelprep}.

\acknowledgments{One of us (MD) would like to thank Gavin Salam for
discussions which helped initiate this study.}

\appendix

\section{Numerical result for the non-inclusive correction}

In this section we evaluate the integral relevant to computing the
non-inclusive correction. Following ref. \cite{Milan1}, one can
write eq. \eqref{eq:mastni} as:
\begin{equation}
\langle \delta p_t \rangle^{ni} = \langle \delta p_t \rangle^{0}
r_{ni},
\end{equation}
with:
\begin{equation}\label{eq:rT}
r_{ni} = \frac{1}{\pi\beta_0} \int_0^1 \frac{dz}{\sqrt{z(1-z)}}\,\,
\frac{1}{2}\int_{-1}^{1}\frac{du_-}{\sqrt{1-u_-^2}}\int_1^\infty
\frac{du_+}{\sqrt{u_+^2-1}} u_1 u_2 \frac{
\mathcal{M}^2}{zu_1^2+(1-z)u_2^2}\widetilde{\Omega}\,.
\end{equation}
Here we have defined $u_\pm=u_1\pm u_2$ and used eq. (4.14) of ref.
\cite{Milan1}. The squared matrix element $\mathcal{M}^2$ is  given
in eqs. (2.8) and (2.9) of the same paper \cite{Milan1}.

By symmetry we just consider the region of the phase-space of
integration defined by $u_->0$ (i.e. $u_1>u_2$) and multiply the
result at the end by a factor 2. This way the trigger function (eq.
\eqref{eq:trigger}) takes the simpler form ($u_1>u_2$):
\begin{eqnarray}
&&\widetilde{\Omega}(u_1,u_2)=\nonumber\\
&&zu_1+(1-z)u_2-\sqrt{zu_1^2+(1-z)u_2^2}+\nonumber\\
&&-\Psi\left(u_2-\mathrm{max}\left[1,
\sqrt{zu_1^2+(1-z)u_2^2-z(1-z)}\right]\right)+\nonumber\\
&&-z \Psi\left(u_1-\mathrm{max}\left[u_2,
1,\sqrt{zu_1^2+(1-z)u_2^2-z(1-z)}\right]\right) \Theta\left(
u_2-\mathrm{min}\left[z/(1-z),1\right]\right)+\nonumber\\
&&+ (1-z) \Psi\left( \sqrt{zu_1^2+(1-z) u_2^2-z(1-z)}-
\mathrm{max}\left[u_2,1\right]\right)\Theta \left(u_2-\mathrm{min}
\left[z/(1-z),1\right]\right),\nonumber\\
\end{eqnarray}
where we made use of the easily-derived inequality
$u_2<\sqrt{zu_1^2+(1-z)u_2^2}< u_1$, for $0<z<1$, and hence
$\sqrt{zu_1^2+(1-z)u_2^2-z(1-z)}<u_1$.

To further simplify the trigger function one needs to consider in
detail the relative size of the various components of the functions
$\Psi$ above. To this end we proceed in the following way.

\subsection{Simplification of the trigger function}

We consider all possible permutations of the relative size of
$\sqrt{zu_1^2+(1-z)u_2^2-z(1-z)}$, $u_1$, $u_2$, and 1. Thus we
discuss the following scenarios:

\begin{itemize}
\item $1<u_2<\sqrt{zu_1^2+(1-z)u_2^2-z(1-z)}<u_1$ or
$1<\sqrt{zu_1^2+(1-z)u_2^2-z(1-z)}<u_2<u_1$. We call the
corresponding region of the phase-space ``A'', thus:
\begin{equation}
\widetilde{\Omega}_A(u_1,u_2)=\sqrt{zu_1^2+(1-z)u_2^2-z(1-z)}
-\sqrt{zu_1^2+(1-z)u_2^2}\,.
\end{equation}
\item $\sqrt{zu_1^2+(1-z)u_2^2-z(1-z)}<1<u_2<u_1$. We label the
corresponding region ``$\widetilde{A}$''. Here we have:
\begin{equation}
\widetilde{\Omega}_{\widetilde{A}}(u_1,u_2)=1-\sqrt{zu_1^2+(1-z)u_2^2}
\,.
\end{equation}
\item $u_2<1<\sqrt{zu_1^2+(1-z)u_2^2-z(1-z)}<u_1$. We name this region
``B''. In this case we have:
\begin{multline}
\widetilde{\Omega}_B(u_1,u_2)=zu_1+(1-z)u_2-\sqrt{zu_1^2+
(1-z)u_2^2}+\\
+\left[ \sqrt{zu_1^2+(1-z)u_2^2-z(1-z)}-zu_1-(1-z)\right]
\Theta\left(u_2-z/(1-z)\right)\Theta(1-2z).
\end{multline}
\item $u_2<\sqrt{zu_1^2+(1-z)u_2^2-z(1-z)}<1<u_1$ or
$\sqrt{zu_1^2+(1-z)u_2^2-z(1-z)}<u_2<1<u_1$.  We call this region
``C'', hence:
\begin{multline}
\widetilde{\Omega}_C(u_1,u_2)=zu_1+(1-z)u_2-\sqrt{zu_1^2+(1-z)u_2^2}
+\\-z\left(u_1-1\right)\Theta\left(u_2-z/(1-z)\right) \Theta(1-2z).
\end{multline}
\item $u_2<\sqrt{zu_1^2+(1-z)u_2^2-z(1-z)}<u_1<1$ or
$\sqrt{zu_1^2+(1-z)u_2^2-z(1-z)}<u_2<u_1<1$. We name this region
``D'', where:
\begin{equation}
\widetilde{\Omega}_D(u_1,u_2)=zu_1+(1-z)u_2-
\sqrt{zu_1^2+(1-z)u_2^2}\,.
\end{equation}
\end{itemize}

We summarise the above results for the trigger function in table
\ref{tab:trigger} for which we define $\Omega_m = zu_1+(1-z)u_2
-\sqrt{zu_1^2 +(1-z)u_2^2}$, which is just the non-inclusive trigger
function for event shapes and the anti-$k_t$ algorithm. We also show
in fig. \ref{fig:phase} the $(u_1,u_2)$ phase-space and the relevant
regions of integration. We note that the overall region of
integration is $u_+>1$ and $-1<u_-<1$ \cite{Milan1}, but we are only
considering the region $0<u_-<1$ as we stated before.
\TABLE[ht]{
\centering
\begin{tabular}{|c|c|}
\hline \textbf{Region} & $\widetilde{\Omega}(u_1,u_2)$  \\
\hline A & $\sqrt{zu_1^2+(1-z)u_2^2-z(1-z)}-
\sqrt{zu_1^2+(1-z)u_2^2}$ \\
&$=\Omega_m+\sqrt{zu_1^2+(1-z)u_2^2-z(1-z)}-zu_1-(1-z)u_2$\\
\hline $\widetilde{\mathrm{A}}$ & $1-\sqrt{zu_1^2+(1-z)u_2^2}
 =\Omega_m +1-zu_1-(1-z)u_2$ \\
\hline B & $\Omega_m+\left[\sqrt{zu_1^2+(1-z)u_2^2-z(1-z)}-zu_1
-(1-z)\right]\times$ \\
& $\qquad\qquad\times\Theta[u_2-z/(1-z)]\Theta(1/2-z)$ \\
\hline C & $\Omega_m- z
\left(u_1-1\right)\Theta[u_2-z/(1-z)]\Theta(1/2-z)$ \\
\hline D & $\Omega_m$\\ \hline
\end{tabular}
\caption{\label{tab:trigger} The trigger function in different
regions of the phase-space. See fig. \ref{fig:phase}.}}
\FIGURE[ht]{\parbox{0.6\textwidth}{ \setlength{\unitlength}{2.1cm}
\begin{center}
\begin{picture}(1,3)
\linethickness{0.2mm} \put(-0.5,0){\vector(1,0){3}}
\put(0,-0.5){\vector(0,1){3}} \put(-0.5,-0.5){\vector(1,1){2.5}}
\put(0.5,-0.5){\vector(-1,1){1}} \put(-0.2,2){$u_1$}
\put(2,-0.2){$u_2$} \put(2,2.2){$u_+$} \put(-0.46,0.5){$u_-$}
\thicklines \put(0,1){\line(1,1){1.5}}\put(1,0){\line(1,1){1.5}}
\put(1,0){\line(-1,1){1}} \put(1,0){\line(0,1){2}}
\put(0,1){\line(1,0){2}} \put(1.35,1.8){A}
\put(1.01,1.15){$\widetilde{\mathrm{A}}$} \put(0.48,0.7){D}
\qbezier(1.2,1.2)(1,1.6)(0.7,1.7) \put(0.85,1.7){B} \put(0.6,1.3)
{C} \linethickness{0.5mm} \put(0,0){\vector(0,1){1}}
\put(0,0){\vector(1,0){1}} \linethickness{1mm}
\put(0,0){\vector(-1,1){0.5}} \put(0,0){\vector(1,1){0.5}}
\end{picture}\newline\newline
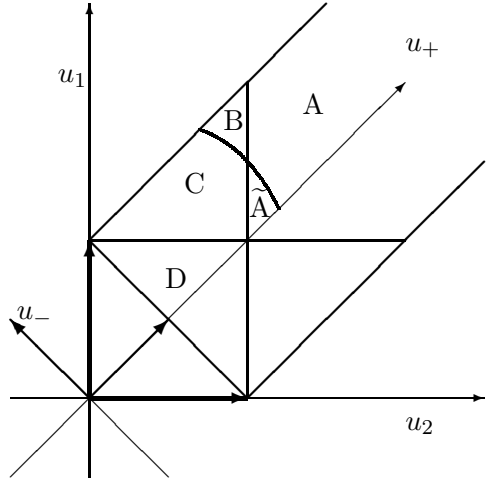
\caption{\label{fig:phase} The $(u_1,u_2)$ phase-space of
integration for eq. \eqref{eq:rT}. See table \ref{tab:trigger} for
the trigger functions $\widetilde{\Omega}$ in various regions. Here
the ellipse (curved line) defines the boundary between regions
governed by the relative size of $z u_1^2+(1-z) u_2^2 -z(1-z)$ and
1. This ellipse ranges from being a straight line parallel to $u_1$
axis (when $z=0$) or $u_2$ axis (when $z=1$) to a circle of radius
$\sqrt{5/2}$ (when $z=1/2$).}
\end{center}
}}

\subsection{Convergence issue}

Having simplified the trigger function we now check the convergence
of the integral \eqref{eq:rT}.

First of all we discuss the limit $u_+\rightarrow \infty$. The
trigger function in this limit (corresponding to region A) is given
by:
\begin{eqnarray}
&&\widetilde{\Omega}_A=\sqrt{z u_1^2+(1-z)u_2^2-z(1-z)}-\sqrt{z
u_1^2+(1-z)
u_2^2}\nonumber\\
&&= -\frac{2z(1-z)}{\sqrt{(u_++(2z-1)u_-)^2+4z(1-z)(u_-^2-1)}+
\sqrt{(u_++(2z-1)u_-)^2+4z(1-z)u_-^2}}\nonumber\\
&&\stackrel{u_+\to\infty}{\longrightarrow} -\frac{z(1-z)}{u_+}\,,
\end{eqnarray}
which clearly vanishes as $u_+\to\infty$, thus ensuring convergence
for reasons cited in \cite{Milan1}.

Next we discuss the limit $z(1-z)\to 0$ and verify the convergence
in various regions of the $(u_1,u_2)$ phase-space. Dealing first
with region A we notice that when $z(1-z)\to 0$ the trigger function
has the behaviour:
\begin{eqnarray}
\label{eq:Atild} \widetilde{\Omega}_A &\stackrel{z(1-z)\to
0}{\longrightarrow}& -\frac{z(1-z)}{u_++(2z-1)u_-}\,,
\end{eqnarray}
which removes the singularity in the matrix element of the form
$1/z(1-z)$  (see ref. \cite{Milan1}). We do not here encounter the
problem of the edge of the phase-space as in \cite{Milan1} ($u_+\to
1$ \emph{and} $u_-\to \pm 1$) since we are away from these corners
of the phase-space in region A, i.e. the denominator in eq.
\eqref{eq:Atild} never approaches zero.

In region $\widetilde{A}$ the trigger function has the following
form:
\begin{eqnarray}
\widetilde{\Omega}_{\widetilde{A}} & = & 1-\sqrt{z
u_1^2+(1-z)u_2^2}\,.
\end{eqnarray}
Here we have the condition $z u_1^2+(1-z)u_2^2-z(1-z)<1$, which can
be written as $-\widetilde{\Omega}_{\widetilde{A}} < z(1-z)/2 +
\mathcal{O}(z^2(1-z)^2)$ when $z(1-z)\to 0$. Thus the trigger
function here removes the singularity in the matrix element of the
form $1/z(1-z)$.

Next we consider region B. The convergence of the integral involving
$\Omega_m$ has been discussed in \cite{Milan1}, so we only check the
convergence of the integral involving the ``correction'' term
$\widetilde{\Omega}_B^{\mathrm{corr}}= \sqrt{zu_1^2+(1-z)u_2^2
-z(1-z)}-zu_1 -(1-z)$, which is present if $z<1/2$ (i.e. we only
worry about $z\to 0$ singularity). In this region we have the
condition $z u_1^2+(1-z)u_2^2-z(1-z)>1$, which can be written as
$-\widetilde{\Omega}_B^{\mathrm{corr}} < (u_1-1) z$, where
$0<u_1-1<1$. Thus we deduce that the trigger function removes the
singularity $1/z$ in the matrix element when $z\to 0$.

Similarly for region C the correction term (which is present if
$z<1/2$) is: $-z (u_1-1)$, which clearly removes the singularity
$1/z$ in the matrix element since $0<u_1-1<1$.

Lastly the integral in region D is obviously convergent since the
trigger function there is equal to $\Omega_m$. Hence we deduce that
the trigger function always removes the $1/z(1-z)$ singularity in
the matrix element.

We note that here we do not specifically have the problem of the
edge of the phase-space which appears in the Milan factor
computation \cite{Milan1} (except for $\Omega_m$, which has been
discussed in the same paper).

\subsection{Numerical result}

Inserting the results for the trigger function from table
\ref{tab:trigger} into eq. \eqref{eq:rT}, taking into account the
appropriate phase-space, we are able to numerically compute the
result for the non-inclusive factor $r_{ni}$. We present here the
numerical result for each region of the phase-space.

We first compute the result for integration over $\Omega_m$ which
appears in all the regions (see table \ref{tab:trigger}). This has
already been done \cite{Milan1} and the result is:
\begin{equation}
r_{ni}^m = \frac{1}{\beta_0} (-1.227C_A+0.365C_A-0.052n_f),
\end{equation}
where after multiplying by the factor 2 (accounting for the fact
that we have considered just $u_->0$) we arrive at the result in
\cite{Milan1}. Next we compute the ``corrections''
region-by--region.

Region D has no corrections, so we begin with region C: the
correction term is $-z\left(u_1-1\right)$, and the phase-space is
$\Theta[u_2-z/(1-z)]\Theta(1/2-z)\Theta(1-zu_1^2-(1-z)u_2^2+z(1-z))
\Theta(u_-^2-(u_+-2)^2)$, with $0<u_-<1$ and $1<u_+<3$. The
numerical result in this case is:
\begin{equation}
r_{ni}^C = \frac{1}{\beta_0} (-0.262C_A+0.072C_A-0.008n_f).
\end{equation}

The correction term for region B is $\sqrt{zu_1^2+(1-z)
u_2^2-z(1-z)}-zu_1 -(1-z)$, and the phase-space is $\Theta[u_2-z/
(1-z) ] \Theta(1/2-z) \Theta (-1+zu_1^2+(1-z) u_2^2-z(1-z))
\Theta(u_-^2-(u_+-2)^2)$, with $0<u_-<1$ and $1<u_+<3$. The
numerical result is:
\begin{equation}
r_{ni}^B = \frac{1}{\beta_0} (-0.043C_A+0.017C_A-0.002n_f).
\end{equation}

For region $\tilde{A}$ the correction is $1-zu_1-(1-z)u_2$ and the
phase-space is $\Theta((u_+-2)^2-u_-^2) \Theta(1-zu_1^2
-(1-z)u_2^2+z(1-z))$, with $2<u_+<1+\sqrt{2}$  and $0<u_-<1$. The
numerical result is:
\begin{equation}
r_{ni}^{\widetilde{A}} = \frac{1}{\beta_0}
(-0.003C_A+0.001C_A-0.000n_f).
\end{equation}

Finally we treat region A. Here the correction reads:
$\sqrt{zu_1^2+(1-z)u_2^2-z(1-z)}-zu_1-(1-z)u_2$, with the
phase-space being $\Theta((u_+-2)^2-u_-^2)
\Theta(-1+zu_1^2+(1-z)u_2^2-z(1-z))$, and $2<u_+<\infty$ and
$0<u_-<1$. The numerical result is:
\begin{equation}
r_{ni}^{A} = \frac{1}{\beta_0} (-0.610C_A+0.154C_A- 0.035n_f).
\end{equation}

Thus the final result for the non-inclusive correction factor
$r_{ni}$, which is $r_{ni}^m+r_{ni}^A+r_{ni}^{
\widetilde{A}}+r_{ni}^B+ r_{ni}^C$ multiplied by a factor 2,
accounting for the fact that we have just considered $u_->0$, is:
\begin{eqnarray}
r_{ni} & = & \frac{2}{\beta_0}
(-2.145C_A+0.610 C_A-0.097n_f)\nonumber\\
&=& \frac{1}{\beta_0} (-3.071C_A-0.194n_f),
\end{eqnarray}
where we have written the result in all the above as a sum of three
terms to show the contributions from the soft gluon, hard gluon and
hard quark matrix elements respectively as in \cite{Milan1}.

Finally we note that $\beta_0 = 11/3 C_A-2/3 n_f = 9 (11)$ for $n_f
= 3(0)$.

\subsection{Comparison between $k_t$ and anti-$k_t$ results}

We present here a comparison between the $r_{ni}$ results for the
$k_t$ and anti-$k_t$ results in various regions of the phase-space.
A factor $2/\beta_0$ is left out: \TABLE{ \centering
\begin{tabular}{|c|c|c|}
\hline Region & anti-$k_t$ & $k_t$  \\ \hline A & $-0.557 C_A+0.123
C_A-0.016 n_f$& $-1.167 C_A+0.277 C_A-0.051
n_f$\\
$\widetilde{A}$ & $-0.000 C_A+0.000 C_A-0.000 n_f$& $-0.003 C_A
+0.001 C_A-0.000 n_f$\\
B & $-0.245 C_A+0.039 C_A-0.004 n_f$& $-0.288 C_A+0.056 C_A-
0.006 n_f$ \\
C & $-0.982 C_A+0.310 C_A-0.017 n_f$& $-1.244 C_A+0.382 C_A-
0.025 n_f$\\
D & $+0.557 C_A-0.107 C_A-0.015 n_f$& $+0.557 C_A-0.107 C_A-
0.015 n_f$\\
\hline Sum & $-1.227 C_A+0.365 C_A-0.052 n_f$& $-2.145 C_A+0.610
C_A-
0.097 n_f$ \\
\hline
\end{tabular}
\caption{Comparison between the non-inclusive coefficient $r_{ni}$
in the $k_t$ and anti-$k_t$ algorithms.}\label{tab:comparison}}

\end{document}